\documentclass[twocolumn,prl,showpacs,preprintnumbers,amsmath,amssymb]{revtex4}
\usepackage[dvipdfm]{graphicx}
\usepackage{dcolumn}
\usepackage{bm}
\usepackage{amssymb}

\begin{document}
\title{Quarter-filled supersolid and solid phases in the extended Bose-Hubbard model}

\author{Kwai-Kong Ng}
\author{Y. C. Chen}
\author{Y. C. Tzeng}
\affiliation{Department of Physics, Tunghai University, Taichung, Taiwan}
\date{\today}
\begin{abstract}
We numerically study the ground state phase diagram of the two-dimensional hard-core Bose-Hubbard model with nearest ($V_1$) and next nearest neighbor ($V_2$) repulsions. In particular, we focus on the quarter-filled phases where one supersolid and two solid phases are observed. Using both canonical and grand canonical quantum Monte Carlo (QMC) methods and a mean field calculation, we confirm the existence of a \textit{commensurate} supersolid at quarter boson filling. The nature of the commensurate supersolid will be discussed. Only one kind of the supersolid phase is found energetically stable despite of the two possible diagonal long range orderings for the solid phase. The competition between the two solid phases manifests as a first order phase transition around $2 V_2\sim V_1$. The change of order parameters as functions of the chemical function is also presented.

\end{abstract}


\maketitle
Bose-Hubbard model has recently attracted a lot of attentions for the possibility of observing the supersolid phase \cite{Penrose} either in optical lattices \cite{Jaksch} or in magnetic systems \cite{Ng1}. The simultaneous breaking of both translational and $U(1)$ gauge symmetry is a delicate state of matter that only recently experimental evidence of possible supersolid phase is provided by the measurement of the non-classical rotational inertia (NCRI) \cite{Andreev} in rotating solid $^{4}$He \cite{Kim}. However, the observed NCRI may be attributed to the superflow in between microcrystal interfaces and the issue is still largely controversial \cite{Burovski}. Thanks to the technological advance in trapping atoms or even molecules at very low temperature in optical lattices, it provides an ideal testing ground for the searching of the  supersolid phase. Hubbard model of hard-core boson in the frustrated triangular lattice \cite{Wessel1} and soft-core boson \cite{Sengupta} in square lattice are among the possible candidates. Besides the optical lattice experiments, various magnetic systems has been suggested \cite{Ng1} to be candidates of the realization of spin supersolid in some carefully chosen parameter regime. In the case of spin $S=\frac{1}{2}$, where the system is equivalent to the hard-core Bose-Hubbard model, the spin supersolid represents the state of spatial modulation of the in-plane spin projection. 

Hard-core Bose-Hubbard model with only nearest neighbor (nn) interaction on a square lattice has ground states of superfluid ordering and checkboard solid ordering. On the other hand, sufficiently large next nearest neighbor (nnn) interaction leads the checkerboard ordering replaced by a striped ordering which can coexist with a superfluid to form a striped supersolid around half-filling \cite{Batrouni}. Recently we revisited the model \cite{Ng2} and found a new solid phase with 'star' ordering (Fig. \ref{starsolid}a) at quarter-filling. Interestingly, this star solid (A-phase) ordering can also coexist with the superfluid ordering to form a star supersolid phase at and around quarter-filling. It has also been suggested \cite{Dang} that another solid (B-phase) order can be stabilized when $2 V_2>V_1$. However, whether the supersolid ground state with this  solid ordering is stable is unclear and the complete ground state phase diagram including the B-phase solid is still absent. These are the questions we attempt to address in this work. Furthermore, there are concerns about the star supersolid phase at exactly quarter-filling for it contradicts the belief that the existence of commensurate supersolid  \cite{Dang,Chen} is impossible. We will show in this paper that the quarter-filled star supersolid is commensurate but agrees with the notion of vacancy supersolid.

Specifically, we study the extended Bose-Hubbard model on a 2D square lattice with the
Hamiltonian
\begin{equation}
H_b =  -t\sum_{i,j}^{nn} (b^\dagger_i b_j+b_i b^\dagger_j) + V_1\sum_{i,j}^{nn} n_i n_j + V_2\sum_{i,j} ^{nnn}n_i n_j - \mu \sum_{i} n_i
\end{equation}
where $b (b^\dagger)$ is the boson destruction (creation) operator
and $\sum^{nn} (\sum^{nnn})$ sums over the (next) nearest neighboring sites. For convenience we fix the energy scale by setting $t=1$ throughout this paper. By taking the transformation $b^\dagger \rightarrow S^\dagger$ and $n \rightarrow S^z+\frac{1}{2}$, this Hamiltonian can be mapped to a spin XXZ model with nn and nnn exchange couplings under a magnetic field $h=\mu -\frac{z}{2}( V_1 +V_2)$ ($z$ is the coordination number). At half-filling, the ground state of the Hamiltonian $H_b$ is a checkerboard solid (characterized by wave vector \textbf{Q}=($\pi,\pi$)) for strong nn coupling $V_1$, or a striped solid (characterized by wave vector \textbf{Q}=$(\pi,0)$ or (0,$\pi$))
for strong nnn coupling $V_2$ \cite{Batrouni}. For $V_1 \sim 2 V_2$, quantum frustration destabilizes both solid orders and leads to a uniform superfluid phase. No supersolid phase is found stable at half-filling. Away from half-filling, however, a striped supersolid is observed for dominating nnn interaction $2 V_2 > V_1$, while the checkerboard supersolid phase is still unstable against phase separation for large nn interaction $V_1 > 2 V_2$. It has been argued that the motion of domain walls reduces the ground state energy so that the checkerboard supersolid is energetically unstable \cite{Sengupta}.

Further away from half-filling, new types of solid in the vicinity of quarter-filling, as well as three-quarter-filling, are found very recently \cite{Ng2,Dang,Chen} as mentioned above. The A-phase quarter-filled solid (see Fig. \ref{starsolid}) has finite structure factor $S(\textbf{Q})/N=\sum_{ij}\langle n_i n_j e^{i \textbf{Q}\textbf{r}_{ij}}\rangle/N^2$ at wavevectors $\textbf{Q}_0=(\pi,\pi)$, $(\pi,0)$ and $(0,\pi)$, which implies a star-like occupation pattern. Moreover, doping the star solid with extra bosons yields a star supersolid via a second order phase transition. The formation of domain walls is no longer energetically favorable and hence the star supersolid is stable upon doping instead of phase separation. More interestingly, this star supersolid persists even at exact quarter-filling for wide range of parameters $V_1$ and $V_2$ (see Fig. \ref{phase}). This result seems to contradict a recent proof \cite{Prokofev} that the necessary condition for supersolidity is to have zero-point vacancies or defects and no commensurate supersolid is possible. Ref. \cite{Dang,Chen} claim they do not observe a quarter-filled star supersolid in their QMC data. Unfortunately, no data of the structure factor and superfluidity are presented in the right $V_1$, $V_2$ parameter regime where we observed the quarter-filled supersolid. One natural question arises about the discrepancy is whether a canonical approach, used in ref. \cite{Chen} where particle number is fixed, leads to different results. In order to clarify the issue, we provide further evidence using Green's Function Monte Carlo (GFMC) to support the existence of quarter-filled supersolid.

\section{Stochastic series expansion}
In order to clarify the issue of the presence of the SS phase at $n=0.25$, we numerically study the model with both the grand canonical and canonical approaches. The  grand canonical calculation is carried out using the standard stochastic series expansion (SSE) Monte Carlo method implemented with directed loop algorithm \cite{Sandvik}. While SSE works on grand canonical ensemble, in order to fix the density $n$ we scan the chemical potential $\mu$ to find an average density $\langle n\rangle=0.25$ with an uncertainty less than 0.01. It turns out, as shown below, this method generates the same result as the GFMC. In SSE, the superfluidity, given by $\rho_{x(y)}=\langle W_{x(y)}^2\rangle/4\beta t$, is computed by measuring the winding number fluctuation as usual.

\section{Green's function Monte Carlo}

The GFMC starts with a Jastrow variational wave function
which is defined by applying the density Jastrow factor to a state
with all the bosons condensed into the $q=0$ state
\begin{equation}\label{wfj}
|\Psi_J \rangle= \exp \left \{ -\frac{1}{2} \sum_{i j} v_{i,j}
n_in_j \right \} |\Phi_0 \rangle,
\end{equation}
where $|\Phi_0 \rangle=(\sum_i b^\dagger_i)^N |0 \rangle$ is the
non-interacting boson ground state with $N$ particles,
 and $v_{i,j}$ are parameters that can be optimized to
minimize the variational energy \cite{sorella}. In order to take
the hard-core constraint into account, configurations with more
than one boson on a single site are projected out from Eq.
\ref{wfj}. The wave function $ |\Psi_J \rangle$ in Eq. \ref{wfj} was shown to be able
to turn a non-interacting bosonic state into a Mott insulator if a
long-range Jastrow factor is included. In our recent work, we have
also shown that the supersolid and solid phases can also be
described in the same wave function. However, the number of the
variational parameters in $v_{i,j}$ grows exponentially with the
lattice size which costs a lot of computation time for an
optimized wave function. Instead of including parameters of all
range, we use a Gutzwiller projection factor $\exp \left \{ \sum_{i\in A} g n_i \right \}$
 to enhance the diagonal order. Here $g$ is a variational parameter which controls the diagonal order specified by the sublattice $A$.
 Obviously, this factor can stabilize the solid and gives 
reasonably good trial energy when $V_1$ and $V_2$ are large. We
found that low variational energy can be acquired for the A-type
supersoild and solid without the projection factor. On the other hand,
relative stable wave function is found by including the projection factor
for the B-type solid.

\begin{figure}
\includegraphics[width=9cm]{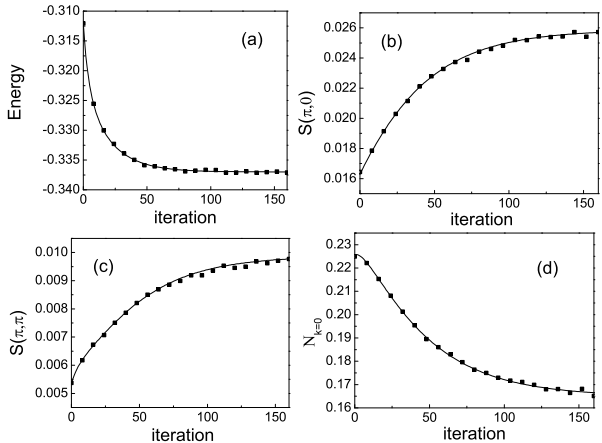}
\caption{GFMC results (solid symbols) of (a) energy, the structure factor of
wavevectors (b) $(\pi,0)$ and (c) $(\pi,\pi)$, and (d) condensate density
$N_{k=0}$(square) as functions of iteration. Solid lines are exact
results without using Monte Carlo. Here $V_1=2$, $V_2=4.5$,
lattice size is $6\times6$ and $n=0.25$. The guiding function are
the same as trial wave function with $v_{1,0}=0.9$, $v_{1,1}=0.83$. 4000 walkers are used in the GFMC calculation.}
\label{EDV245optg}
\end{figure}

In order to investigate the exact ground state properties, Green's
function Monte Carlo method is employed to improve the variational
results. In this work, we choose multi-walkers stochastic
reconfiguration method to prevent the simulation from blow-up or
dead-ends in the large power limit. To benchmark our method we
compare our GFMC data (points) with the exact results (lines)
without using Monte Carlo for an $6\times6$ lattice in Fig.
\ref{EDV245optg}. We can see that the GFMC results are consistent
with the exact ones. The underestimation of the diagonal order in
the variational wave function is corrected as the iteration
increases. We have verified that the same ground state properties
can also be obtained with the wave function with diagonal order
($g\ne0$).

\begin{figure}
\includegraphics[width=9cm]{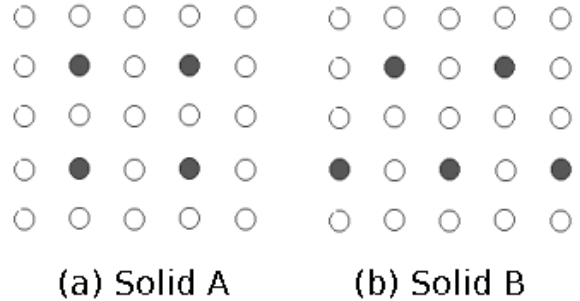}
\caption{Classical lattice structures of star solid (a) A and (b) B phase.}
\label{starsolid}
\end{figure}

\begin{figure}
\includegraphics[width=9cm]{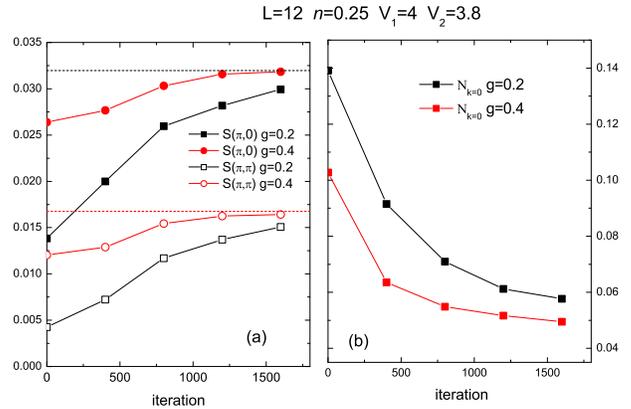}
\caption{(Color online)(a) The GFMC results of the structure factor
of wavevectors $(\pi,0)$, $(\pi,\pi)$, and (b) condensate density
$N_{k=0}$(square) as functions of iteration. Dash lines are SSE
results. $V_1=4$, $V_2=3.8$, lattice size is $12\times12$ and
$n=0.25$.} \label{GFMC-L12V438}
\end{figure}

\begin{figure}
\includegraphics[width=9cm]{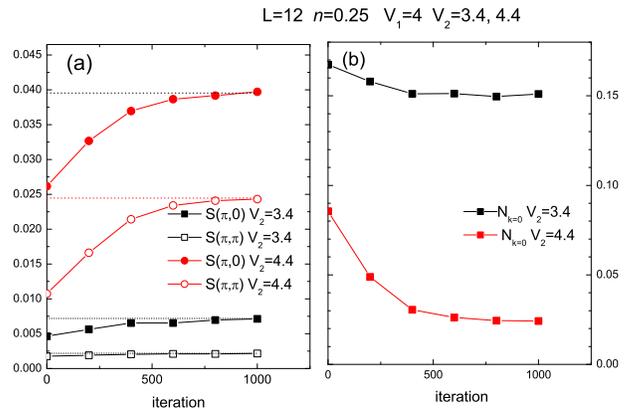}
\caption{(Color online)(a) The GFMC results of the structure factor
of wavevectors $(\pi,0)$, $(\pi,\pi)$, and (b) condensate density
$N_{k=0}$(square) as functions of iteration. Dash lines are SSE
results. Here $V_1=4$, $V_2$=3.4 and 4.4. Lattice size is
$12\times12$ and $n=0.25$.} \label{GFMC-L12V434n44}
\end{figure}

Now we present the GFMC results for larger lattice. In Fig.
\ref{GFMC-L12V438} we show the structure factors and condensate $N_{k=0}=b_{k=0}^{\dag}b_{k=0}$ as
a function of iteration. The trial wave function with A-type
diagonal order is used as the trial wave function. In order to
check that the ground state can be reached regardless of the
choice of $g$, we present results of $g=0.2$ and $g=0.4$. The wave
function is optimized for all parameters before GFMC is applied.
The corresponding data obtained by SSE (dash lines) are also shown
for comparison. As we can see that although the optimized wave
function overestimated the structure factor and underestimated the
condensate density, consistent results with SSE are obtained as the
iteration increases. Fig. \ref{GFMC-L12V434n44} are similar
calculations for $V_2$=3.4 and 4.4 corresponding to superfluid and
A-type solid phases, respectively. The data clearly show the
convergence of the GFMC approach and consistent with the SSE
results.

\section{Mean field theory}

To further investigate the effect of quantum fluctuation of the model, we also obtain the ground state phase diagram using a simple mean field approach. A mean field wavefunction 

\begin{equation}
|\Psi\rangle_{MF}=\prod_{i,\alpha}( u_i |1\rangle_{i,\alpha} + v_i |0\rangle_{i,\alpha})
\end{equation}
is given to represent the superfluid, star solid and star supersolid phases. Here $\alpha$ denotes the index of a 2x2 unit cell while $i=1,2,3,4$ is the sublattice index inside the cell. $|1\rangle_{i,\alpha}$ ($|0\rangle_{i,\alpha}$) is the occupied (empty) Fock state at $i$-th site of the unit cell $\alpha$ while $u_i (v_i)$ is the corresponding variational parameter. The energy of $|\Psi\rangle_{MF}$ is then minimized to obtain the ground state wavefunction. This MF wavefunction successfully predicts the existence of quarter-filled star supersolid as well as the superfluid and star solid as shown in the inset of Fig.\ref{phase}. 

\section{Quarter-filled supersolid}

\begin{figure}
\includegraphics[width=7cm]{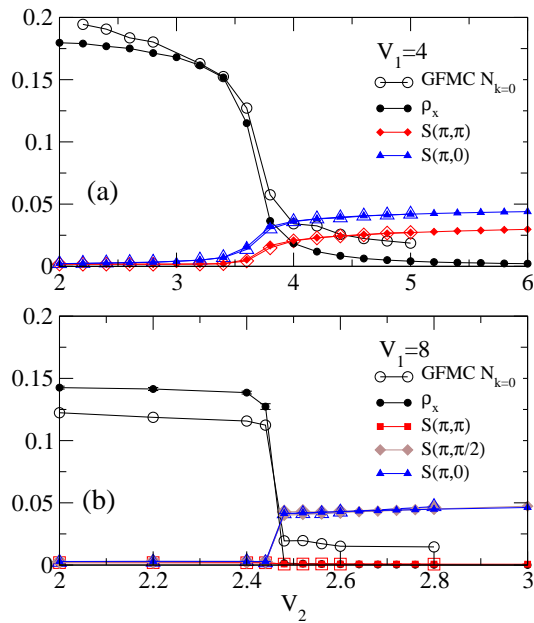}
\caption{(color online) Comparison of data obtained from the SSE (filled symbols) and GFMC (open symbols) for lattice size $L=12$. Note that GFMC measures condensate density $N_{k=0}$ while SSE measures superfluidity $\rho$ for the signature of superfluid phase. Temperature used in the SSE is T/t=0.02.}
\label{12x12}
\end{figure}

\begin{figure}
\includegraphics[width=7cm]{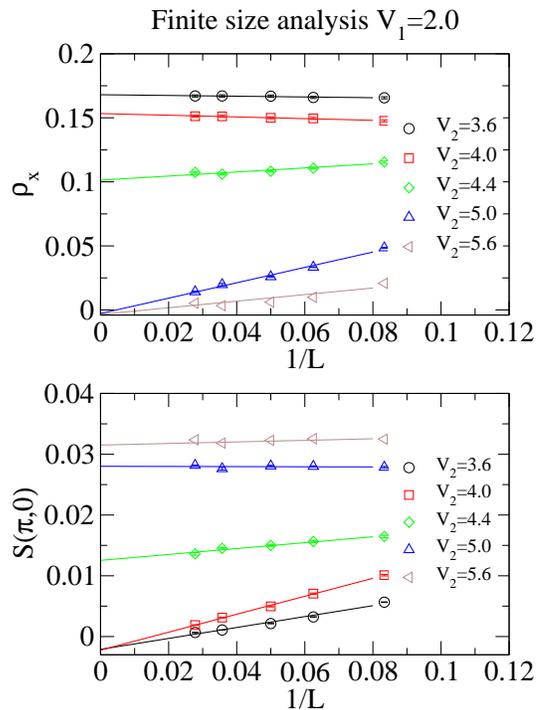}
\caption{(color online) Finite size analysis of the order parameters $\rho_x$ and $S(\pi,0)$ from SSE for $V_1=2.0$.}
\label{fs}
\end{figure}

\begin{figure}
\includegraphics[width=7cm]{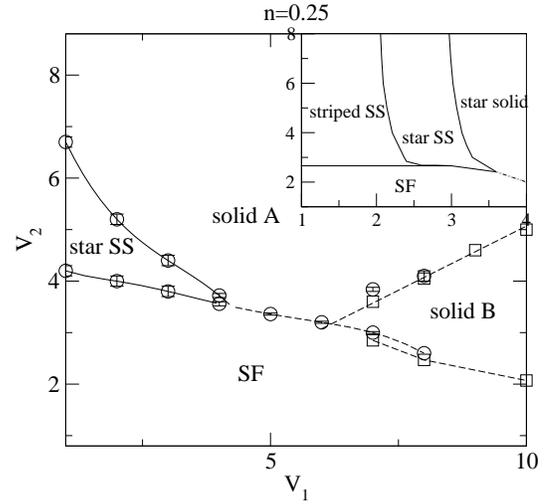}
\caption{Phase diagram of $n=0.25$. The lines are guides to the eyes. Solid (dotted) lines represent second (first) order phase transition boundaries. Black circles (squares) denote the data obtained from the SSE (GFMC) for L=28 (L=16). The inset shows the MF result. It is noted that star solid A and B phase cannot be distinguished in the MF level.}
\label{phase}
\end{figure}

Fig.\ref{12x12} shows the result of SSE and GFMC at $V_1=4$ and $V_1=8$ for small lattice size. The agreement between both approaches is clear and verifies that our attempt to fix the boson density $n$ to 0.25 in the grand canonical SSE does not lead to any measurable discrepancy of the physical quantities we are interested in with the canonical GFMC. The coexistence of superfluid order and star crystal structure for $3.5\lesssim V_2 \lesssim 4.0$ in Fig.\ref{12x12} (a) clearly signals the supersolid phase at quarter-filling. The uniform superfluid develops spatial modulation, i.e. becomes a supersolid, continuously as $V_2$ increases and gradually loses its superfluidity at the same time until it eventually becomes a star solid. To demonstrate that the supersolid ground state survives at larger lattice size and at thermodynamic limit, a finite size analysis is carried out in Fig. \ref{fs} for $V_1=2$. $S(\pi,0)$ scales to zero in the superfluid phase whereas $\rho_x$ scales to zero in the solid phase. Only in the supersolid phase ($V_2=4.4$ in Fig. \ref{fs}) that both $S(\pi,0)$ and $\rho_x$ scales to finite values. It is remarkable that simple MF theory also correctly predicts the existence of quarter-filled SS and quantum fluctuation does not destroy the long-range orders in the SS phase, in contrast to the case of Kagome lattice \cite{Isakov}. Except that the solid A and B phase are indistinguishable in the MF level, MF theory successfully reproduce all phases at quarter-filling as shown in the inset of Fig. \ref{phase}. 

\begin{figure}
\includegraphics[width=7cm]{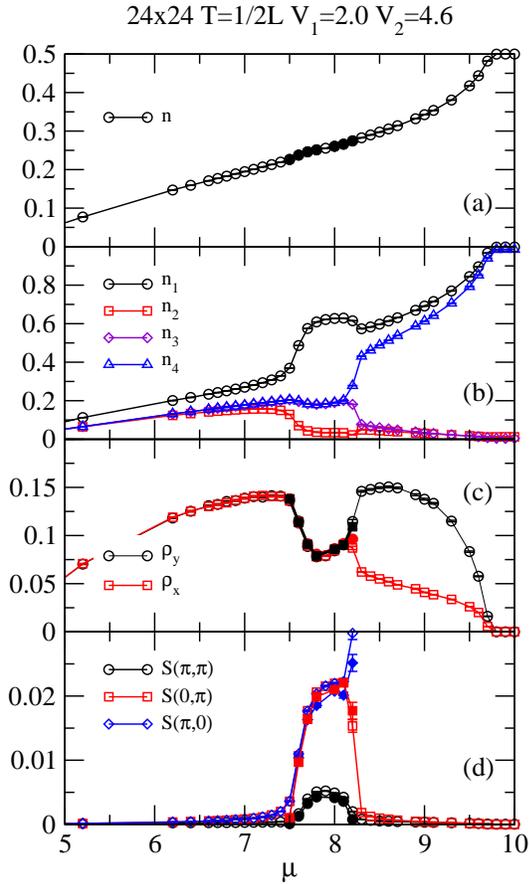}
\caption{(color online) SSE result of the (a) average density $n$, (b) sublattice densities, (c) superfluidity and (d) structure factors as functions of chemical potential $\mu$ for $V_1=2.0$ and $V_2=4.6$. Open (filled) symbols denote the result for a lattice of size 24x24 (48x48).}
\label{vary_mu}
\end{figure}

In the first glance, the existence of a supersolid at \textit{commensurate} density contradicts the notion of vacancy supersolid. Prokof'ev and Svistunov \cite{Prokofev} have proved that  superfluidity has zero probability to occur in commensurate solids in nature, or in other words, the necessary condition for the supersolidity is the present of vacancies or defects. This is due to the asymmetry between vacancies and interstitials. However, this result, as admitted in their paper, does not apply to systems with explicitly broken translation symmetry, such as in lattice models where commensurability is enhanced by hand. 
In our system, the commensurability of quarter-filled supersolid can always be ensured by adjusting the chemical potential provided there is no phase separation. Vacancies and interstitials (means bosons at the sublattice sites) arise from quantum fluctuations do not form bound pairs and are free to move away from each other that, therefore, leads to superflow. The interstitial-vacancy symmetry is generally absent in nature but is preserved in this case by the external potential that fix bosons to locate only at the lattice points.  Based on the measured structure factors $S(\pi,\pi)$,  $S(\pi,0)$ , $S(0,\pi)$ and boson density $n$, one can easily deduce the boson densities on each sublattices. Fig.\ref{vary_mu} shows that bosons do not localized at only one sublattice but have finite occupation in all four sublattices. The data indicates that there are more than 30 percent of vacancies in the quarter-filled supersolid and in consistent with the picture of vacancy supersolid. 

To gain more insight of the quarter-filled supersolid phase, we plot in Fig.\ref{vary_mu} the order parameters as functions of the chemical potential $\mu$. Starting from small density, the ground state is an uniform superfluid in which all sublattice densities are equal. While increasing $\mu$ until quarter-filled, the system undergoes a second order phase transition to a star SS. Note that although there is a small dip of superfluidity in the SS state, it does not reduce to zero even at $n=0.25$. There is no any indication that $n=0.25$ is a special density that acquires particular treatment like the canonical calculation. Furthermore, no noticeable change is observed when doubling the lattice size to 48x48 (filled symbols in Fig.\ref{vary_mu}). Doping more bosons will destabilize the star SS phase because of the nnn repulsion that leads to a striped SS via a discontinuous phase transition between the two different broken translation symmetries. Striped SS is also observable at $n=0.25$ when $V_1$ is small enough. On the other hand, while there are two kinds of star solid (A and B phase), a natural question is whether SS of both kinds exist at or away from quarter-filled . We will address this issue in the following section.

\section{Supersolid A and B phase}

\begin{figure}
\includegraphics[width=7cm]{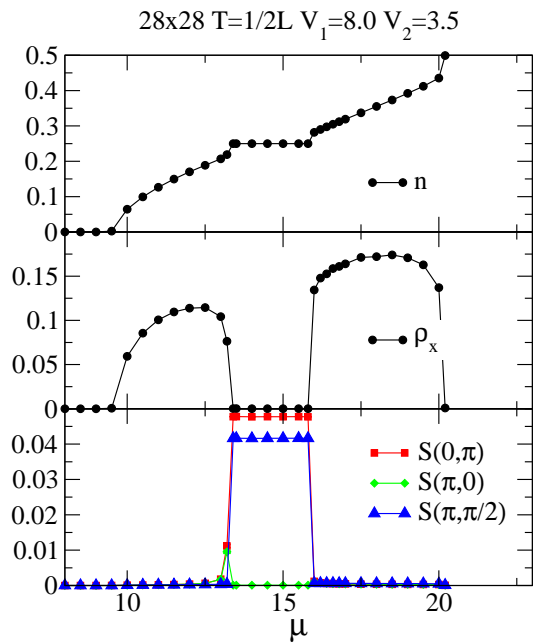}
\caption{(Color online) SSE result for $V_1=8.0$ and $V_2=3.5$. The ground state at $n=0.25$ is a B phase solid. Doping away from $n=0.25$ leads to a first order phase transition to SF phase. No B phase supersolid is found.}
\label{B_mu}
\end{figure}

It is interesting to note that at quarter-filling, the classical ground state of the frustrated Hamiltonian $H$ is highly degenerate. Translating any lines of bosons in Fig. \ref{starsolid}a by one lattice constant of the solid A phase along, say, $x$ direction will create a domain wall with no energy cost. When translating alternate lines of bosons generate a B phase solid. Enormous number of ways to create domain walls implies that the classical ground state has macroscopic degeneracy. This degeneracy, however, is lifted by quantum fluctuation that yields a ground state of A phase if $V_1 < 2 V_2$ or B phase otherwise. This is another typical example of the order by disorder phenomenon\cite{Wessel1}. As discussed in previous sections, it leads to a star SS A phase in a wide parameter range, similar to the scenario of the one-third filling SS in the frustrated triangular lattice \cite{Wessel1}. On the other hand, the existence of star SS B phase has not been clarified. Here we complete the phase diagram for larger $V_1$ where solid B phase can be stabilized at quarter-filling. Fig. \ref{phase} shows that, for $V_1>6$, when increasing from small nnn interaction $V_2$, the SF phase changes to the solid B phase via a first order phase transition without passing an intermediate SS B phase. Star solid A phase emerges when further increasing $V_2$ until $2 V_2 \gtrsim V_1$. Furthermore we do not observe any SS B phase away from $n=0.25$ as shown in Fig. \ref{B_mu} at a representative $V_1=8.0$ and $V_2=3.5$. Instead there is a first order phase transition from the gapped solid B phase to a uniform SF phase and implies that the SS of B phase symmetry is unstable towards phase separation. This can be understood by the simple argument of domain formation as discussed in the case of hardcore checkerboard SS \cite{Sengupta}. Addition holes (bosons) in the quarter-filled solid phase B tend to line up to form a domain wall in which the gain in kinetic energy is larger than that of isolated holes (bosons). The SS B phase is therefore unstable towards phase separation. 

\section{Summary}
We have presented a comprehensive numerical study on the extend hardcore Hubbard model, in particular, on the SS phases at the quarter-filled density. Based on results of the SSE, GFMC and MF calculations, we provide convincing evidence for the existence of star SS A phase at exact $n=0.25$, in contrasts to previous study. We clearly show that the SS A phase found at $n=0.25$ is consistent to the notion of vacancy SS. The star SS phase is a consequence of the order by disorder phenomenon by which the ground state degeneracy is lifted. Furthermore, the physical natures the solid A and B phase are also studied. Although classically degenerate, these two states compete with each other through quantum fluctuations and the final stability depends on the competition between nn ($V_1$) and nnn ($V_2$) interactions. We present a complete phase diagram including also the solid B phase. The SS B phase on the other hand is found unstable towards phase separation due to the kinetic energy gain of domain formation. 

\begin{acknowledgments}
We are grateful to M.F. Yang for simulating discussion. K.K.N. acknowledges the financial support by the NSC (R.O.C.), grant no. NSC 97-2112-M-029-003-MY3 and NSC 95-2112-M-029-010-MY2.
\end{acknowledgments}

\end{document}